\def\plainformat{}
\documentclass{article}

\usepackage{amsmath}
\usepackage[super]{nth}

\usepackage{siunitx}

\usepackage{tikz}
\usetikzlibrary{chains,matrix,scopes,positioning,arrows}

\ifx\plainformat\undefined\newcommand{\switchcitation}[1]{\cite{#1}}\else\newcommand{\switchcitation}[1]{~\cite{#1}}\fi

\begin{document}
\ifx\plainformat\undefined\markboth{L.~J.~Gunn, J.~M.~Chappell, A.~Allison, D.~Abbott}{Physical-layer encryption on the public internet}\fi
%
\ifx\plainformat\undefined\catchline{}{}{}{}{}\fi
%
\title{Physical-layer encryption on the public internet: a stochastic approach to the Kish-Sethuraman cipher}
\ifx\plainformat\undefined{
\author{Lachlan J. Gunn}
\author{James M. Chappell}
\author{Andrew Allison}
\author{Derek Abbott}
}\else{\author{Lachlan J. Gunn \and James M. Chappell \and Andrew Allison \and Derek Abbott}}\fi
\ifx\plainformat\undefined{
 	\address{School of Electrical and Electronic Engineering\\
	The University of Adelaide, SA 5005, Australia\\
	lachlan.gunn@adelaide.edu.au}
	\maketitle
}\else{
 	\maketitle
 	\footnotetext[1]{School of Electrical and Electronic Engineering\\
	The University of Adelaide, SA 5005, Australia\\
	lachlan.gunn@adelaide.edu.au}
	\footnotetext[2]{To be presented at HotPI-2013}
}\fi

\ifx\plainformat\undefined{
\begin{history}
\received{Day Month Year}
\revised{Day Month Year}
\end{history}
}\fi

\begin{abstract}
While information-theoretic security is often associated with the one-time pad and quantum key distribution, noisy transport media leave room for classical techniques and even covert operation.  Transit times across the public internet exhibit a degree of randomness, and cannot be determined noiselessly by an eavesdropper. We demonstrate the use of these measurements for information-theoretically secure communication over the public internet.
\ifx\plainformat\undefined\fi
\end{abstract}

\ifx\plainformat\undefined\ccode{PACS numbers: 84.40.Ua, 89.20.Ff, 89.20.Hh, 89.70.Cf, 89.70.Kn}\fi

\section{Introduction}	

Throughout history claims have abounded of supposedly unbreakable codes and
ciphers, however it was not until the \nth{20} century that the mathematical
underpinnings of cryptology gave them any degree of credibility.

The first truly unbreakable cipher was the
one-time-pad\switchcitation{kahn-the-codebreakers}, invented in the United States
in 1918, but independently developed and first put into practice by the
German Foreign Office in the early 1920s.  This scheme, however, is hampered by
the need to distribute a key to the message recipient of the same size as
the message and in perfect secrecy.  This onerous key distribution
arrangement ruled it out for all but the most critical applications.

In recent years BB84, introduced by
Bennett and Brassard\switchcitation{bb84}, provides security by encoding data in one of
two non-orthogonal bases, chosen at random for each photon.  The measurement basis
is also chosen at random.
After the measurement has taken place, the two parties reveal their
chosen bases and discard those bits that were not measured in the correct
manner.  As an eavesdropper does not know in advance which basis was used,
she cannot reliably copy its polarisation state and so
any eavesdropping will manifest itself as an increase in the bit-error rate
(BER). This allows eavesdropping to be detected by the two legitimate parties.

Later, Maurer\switchcitation{maurer-key-agreement} considered a more general case, letting
the sender (`Alice' hereafter), recipient (Bob), and eavesdropper (Eve)
each perform measurements $X$, $Y$, and $Z$ respectively.  He demonstrated that
secure information exchange could occur (subject to some constraints) with an
arbitrary joint distribution $P_{XYZ}$, reopening the door to
information-theoretically secure communication over a classical channel in
some circumstances.

In 2004, Kish and Sethuraman proposed a classical
protocol\switchcitation{kish-sethuraman} based upon commutative one-way
encryption operators.  If the sender and recipient each apply a layer
of encryption to a message, then commutativity allows the sender to reverse her
own operation and so produce a ciphertext with key known only to the recipient.
In order to overcome the information-theoretic limits given by
Maurer\switchcitation{maurer-key-agreement} we have attempted to relax the constraints
of the protocol\switchcitation{kish-sethuraman} by using the random transit times of the internet
as encryption operations---this imperfect channel allows secure
communication without resort to the one-time pad.

\section{Round-trip times as a source of randomness}

The essence of key distribution is to provide two endpoints with a shared secret
that remains unknown to eavesdroppers.  A subtle point is that the endpoints
need not generate a secret and share it themselves, but could instead obtain it
from elsewhere, provided an eavesdropper cannot do the same without error.

\begin{figure}[h]
	\centering
	\includegraphics[width=0.4\linewidth]{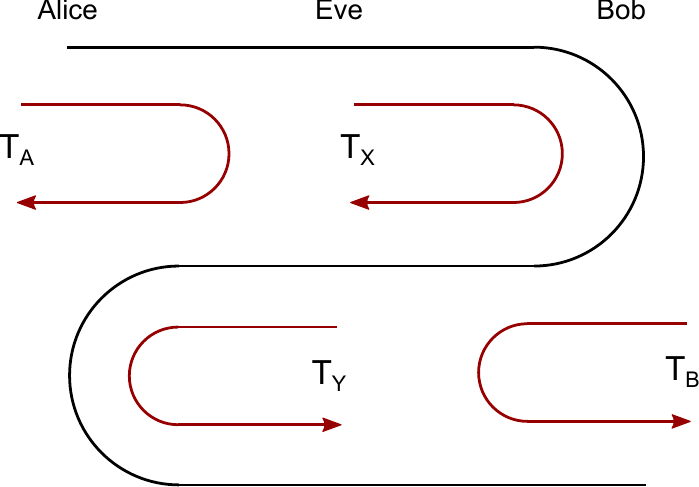}
	\caption{Consecutive round-trip measurements, where Bob's
			response to Alice forms a request for another measurement.  The
			transit time of this intermediate transmission contributes to the 
			measurements of Alice, Bob, and Eve, and so provides a source of
			mutual information.}
	\label{fig:round-trip-diagram}
\end{figure}

One such source of random data is the transit time between two internet-connected
terminals.  If Alice and Bob rally information packets back and forth via the
internet, the time of each
transit is a quantity common to the measurements of both, but measurable only
with the addition of
noise from the return trip (see Figure~\ref{fig:round-trip-diagram}).  An
eavesdropper will suffer the same problem,
however her noise will differ from that of Alice and Bob.  This difference
prevents her from taking advantage of the error correction performed by Alice
and Bob during the information reconciliation\switchcitation{information-reconciliation}
(IR) phase of processing, which discards bits likely to be incorrect, much like
in the quantum protocol.

We propose to extract random bits from the round-trip times by finding their
median and declaring those times greater than the median to be a one, and those
less to be a zero.  With only one bit per round-trip, we avoid the problem that
errors are more likely to fall into
a adjacent quantisation bins and so create correlations between bits.

While the distribution of round-trip times is actually quite
skewed\switchcitation{round-trip-distribution}, we attempt to illustrate the technique
theoretically by assuming transit times to be normally distributed and
computing an upper limit on the key rate.

\subsection{The mutual information rate between endpoints}

Let us denote the three packet transit times from
Figure~\ref{fig:round-trip-diagram} as $T_1$, $T_2$, and $T_3$ respectively.  Then,
$T_A = T_1 + T_2$, $T_B = T_2 + T_3$. Suppose the three
$T_i \sim \mathcal{N}(0,1)$.  Then, as the distribution (and so the channel)
is symmetric, we may calculate the bit error rate as
\begin{align}
	1-P[T_B < 0 | T_A < 0]	&= 1-P[T_2 + T_3 < 0 \; | \; T_1 + T_2 < 0] \\
					&= 1-\int_{-\infty}^{+\infty} 2\phi(t_2) P[T_3 < -t_2 \cap T_1 < -t_2] \; dt_2 \\
					&= 1-\int_{-\infty}^{+\infty} 2\phi(t_2) \Phi^2(-t_2) \; dt_2 \\
					&= 1-\int_0^1 2u^2 \; du \\
					&= \frac{1}{3} ,
\end{align}
where $\phi(t)$ and $\Phi(t)$ are the probability density and cumulative
probability functions respectively of the normal distribution function.  It
should be noted that the derivation above holds for any zero-median symmetric
distribution rather than just for the normal distribution.

This BER of $1/3$ corresponds to a channel capacity of 0.08 bits/measurement,
suggesting that the achievable key rate with this technique may be too low for
direct use as a one-time pad.

\subsection{Limitations}

Despite the allure of information-theoretically secure key agreement without
specialised hardware, this method is not perfect and is necessarily dependent
on the eavesdropper's inability to timestamp packets with perfect accuracy.
This limits its use where an eavesdropper can timestamp
packets on the link directly.  To illustrate this point, we make use of Maurer's upper
bound\switchcitation{maurer-key-agreement} on the secrecy rate, the maximum rate at
which information can be transmitted securely,
\[
S(X;Y || Z) \leq I(X;Y) . \label{eqn:sr-bound}
\]
This states that the rate at which secure communication may take place is
limited to the mutual information of the two endpoints.  The effect of this
statement is that a protocol, no matter how clever, cannot provide
secrecy using only independent random number generators at each end.

In order to demonstrate the relevance of this inequality,
imagine that Eve can timestamp Alice's transmissions without error.  Then, Alice
cannot gain an advantage over Eve, who has the same information as Alice,
and so secrecy is impossible.

Now imagine that we have placed a router between Alice and Bob.  This introduces
some randomness, but Alice could achieve the same effect by simply
adding a random delay to her transmissions; that is to say, it is as though
she used a random unshared key.  As discussed, this cannot form the basis for
a secure system. Therefore, if Eve can measure without
noise, information-theoretic security is not possible.

However, there are many cases in which this is not true.  If an eavesdropper
simply has copies of all traffic forwarded to them (such as provided by
port mirroring), then routing delays and
packet reordering provide the necessary source of noise\switchcitation{portmirroring}.
If a standard PC is acting as a man-in-the-middle, uncertainty in the timing
routines of its operating system provide an additional source of noise.
These factors allow the system to provide security, especially against
unsophisticated eavesdroppers using only commodity network hardware that
does not provide hardware timestamping facilities.

\section{Experimental Round-Trip Measurements}
\begin{figure}[h!]
	\centering
	\includegraphics[width=0.58\linewidth]{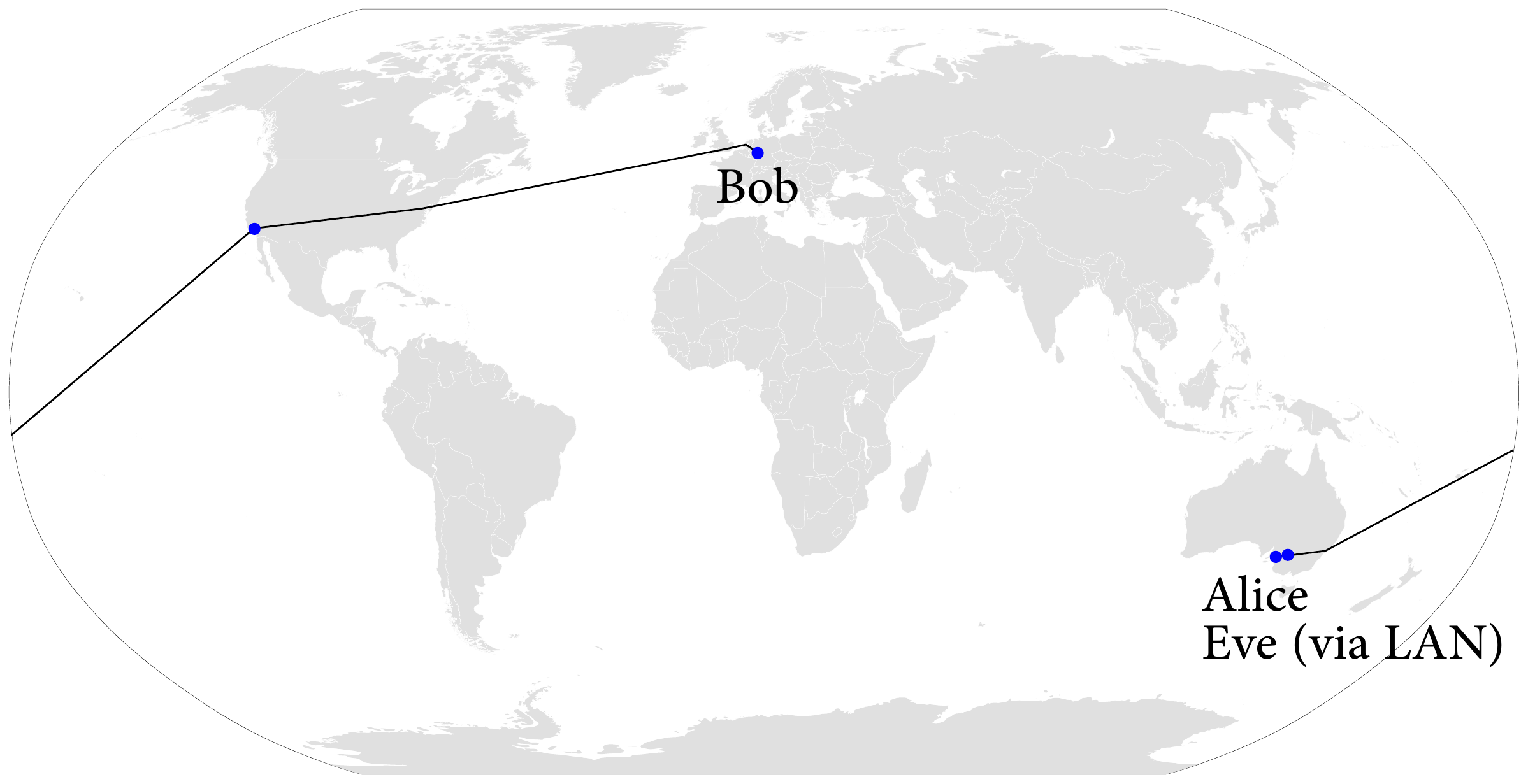}
	\caption{The constructed communications link.  Alice is located in Adelaide, Australia,
			in the same room as Eve, to whom she is connected via the local network.
			The packet is then forwarded to through a relay in
			Los Angeles, and finally to Bob in Frankfurt. Alice sends a packet to Eve,
			who forwards it to Los Angeles, and finally to Bob in Frankfurt.  The
			packet is then returned.  At each step the arrival of the packet is
			timestamped, and the packet finally returned to Alice contains a time of
			arrival for Bob, and two for each intermediate note. The head of the chain in
			Adelaide, representing Alice, sends a packet which is timestamped at
			each of the three other hosts.  The demonstration system described later
			does not transmit timestamps over the network, allowing each node to
			determine only its own round-trip time.}
	\label{fig:chain}
\end{figure}

In order to demonstrate this technique, we constructed a
test system to determine the performance of the method in the presence of an
eavesdropper.   The test system rallied UDP packets back and forth along a
chain of hosts (see Figure~\ref{fig:chain}),
with the time of each arrival being timestamped.  While the timescales were not
synchronised, this information is sufficient to determine the various round-trip
times.

\begin{figure}[h]
	\centering
	\includegraphics[width=0.6\linewidth]{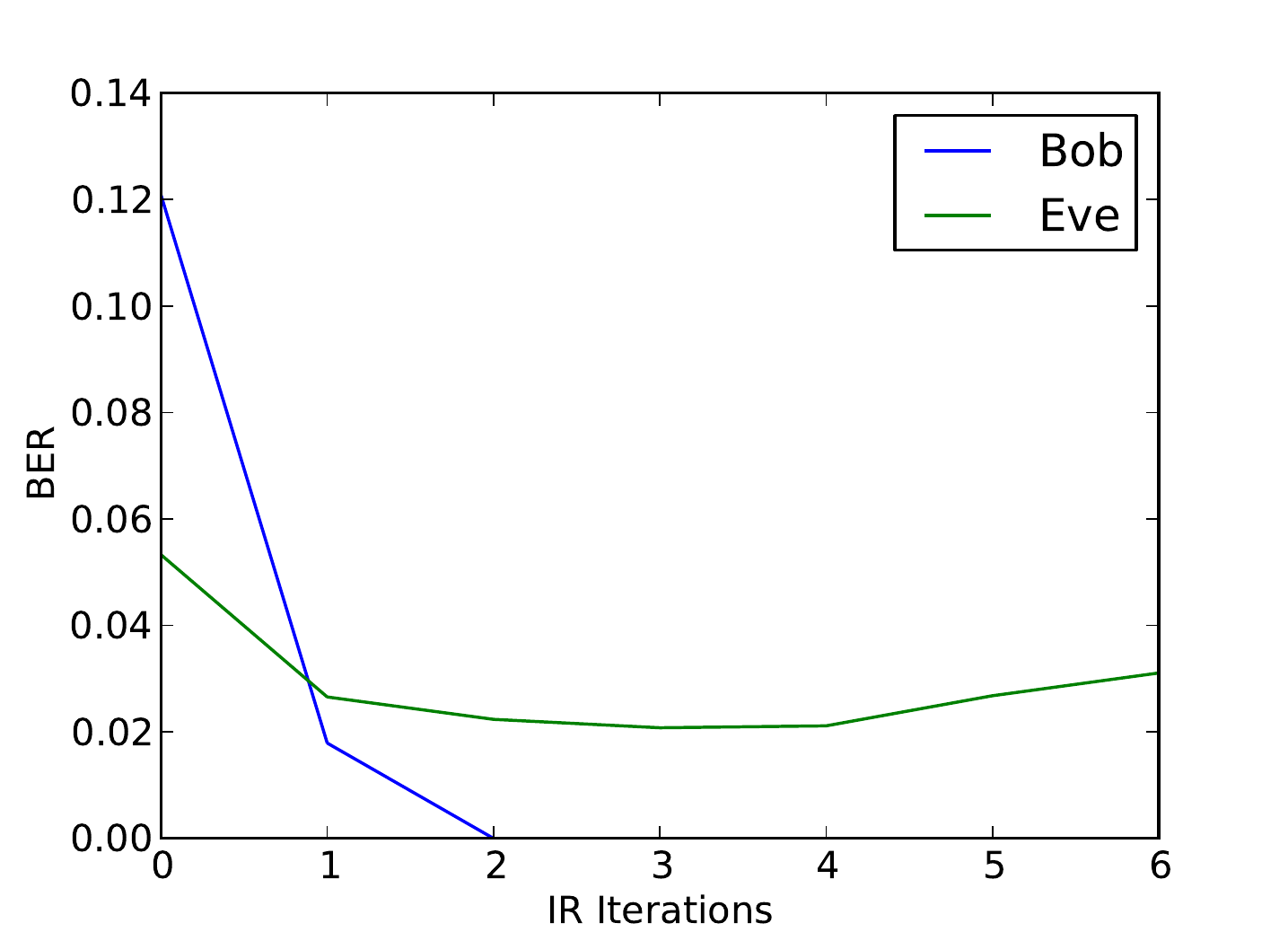}
	\caption{The effect of bit-pair iteration upon the BER between
			Alice/Bob and Alice/Eve.  This measurement used
			approximately 30,000 round-trips AU-US-EU, with the eavesdropper
			chosen to be the node in the same room as the sender.  While
			the BER of the eavesdropper is improved slightly, it is not reduced to
			zero, which is evidence that there is sufficient measurement noise to allow
			secure communication.}
	\label{fig:ir-ber}
\end{figure}

The effect of information reconciliation is shown in Figure~\ref{fig:ir-ber}.
While the BER of the
eavesdropper falls at first, it soon reaches a minimum value of around 2\%.
This demonstrates that some nonzero secrecy rate is achievable via privacy
amplification.

\section{Demonstration System}

We have implemented the described protocol, which has been successfully operated
over the internet.  Round-trip times are measured using UDP packets, whose times
of transmission and receipt are determined using operating system routines.  If
a timeout occurs, due to a dropped packet for instance, the trip is marked as
such and dropped during the reconciliation process. Information reconciliation
is performed using the bit-pair
iteration\switchcitation{maurer-key-agreement} protocol, each endpoint executing
identical code in lock-step (with the exception of a network abstraction
layer).

Parameters for the information reconciliation and privacy amplification are
determined automatically.  A lower bound on eavesdropper BER is given as
a parameter, and so their channel capacity is computed and thus the amount of
information that they hold.  From this, a hashing function is chosen---the sum of
some number of bits modulo two---that will discard sufficient information to
eliminate the eavesdropper's knowledge of the secret key.  As this process will
increase the BER of the legitimate parties also, the target BER for the information reconciliation is reduced
to compensate.

The BER of the channel is estimated
using the error rate of the parity bits.  A $2\sigma$
Agresti-Coull\switchcitation{agresti-coull} confidence interval is constructed and
back-propagated through the binomial probability mass function\switchcitation{larsen-statistics}, yielding
a confidence interval for the
BER of the underlying channel.  Then, the BER of the parity-checked output of
each iteration can be computed recursively in order to determine an interval
containing the required number of IR iterations.  When this interval has been
reduced to two possible values, we take the larger of the values and continue
with the complete reconciliation process.

We succeed in generating keys at a rate of 13 bits/minute over the link shown in Figure~\ref{fig:chain},
the lower bound on the eavesdropper BER set at $10^{-2}$, based on the results
shown in Figure~\ref{fig:ir-ber}.
The \SI{400}{\milli\second} round-trip time makes the
test relatively pessimistic by terrestrial standards, and greater
key rates are potentially achievable across shorter distances.

\section{Conclusion}

We have considered the use of packet timing over the internet for
information-theoretically secure key agreement, and demonstrated the
feasibility of the technique experimentally.  We have 
developed a practical implementation of the method that is capable
of generating shared keys in real-time despite the assumption of an eavesdropper
BER equivalent to a man-in-the-middle attack from within the local network.


\end{document}